\documentclass[useAMS,usenatbib,usegraphicx]{mn2e}
\usepackage{graphicx}

\renewcommand\theequation{\arabic{equation}}
\newcommand{\bea}{\begin{eqnarray}}
\newcommand{\eea}{\end{eqnarray}}

\title[]{High-redshift star formation rate up to $z\sim8.3$ derived from gamma-ray bursts and influence of
background cosmology}
\author[F. Y. Wang and Z. G. Dai] {F. Y. Wang \thanks{fayinwang@nju.edu.cn} and Z. G. Dai\thanks{dzg@nju.edu.cn}\\
 Department of Astronomy, Nanjing University, Nanjing 210093, P.
R. China}
\begin{document}

\maketitle


\begin{abstract}
The high-redshift star formation rate (SFR) is difficult to measure
directly even by modern approaches. Long-duration gamma-ray bursts
(GRBs) can be detected to the edge of the visible universe because
of their high luminorsities. The collapsar model of long gamma-ray
bursts indicates that they may trace the star formation history. So
long gamma-ray bursts may be a useful tool of measuring the
high-redshift SFR. Observations show that long gamma-ray bursts
prefer to form in a low-metallicity environment. We study the
high-redshift SFR up to $z\sim8.3$ considering the \emph{Swift} GRBs
tracing the star formation history and the cosmic metallicity
evolution in different background cosmological models including
$\Lambda$CDM, quintessence, quintessence with a time-varying
equation of state, and brane-world model. We use latest \emph{Swift}
GRBs including two highest-$z$ GRBs, GRB 080913 at $z=6.7$ and GRB
090423 at $z=8.3$. We find that the SFR at $z>4$ shows a steep decay
with a slope of $\sim -5.0$ in $\Lambda$CDM. In the other three
models, the high-redshift SFR is slightly different from
$\Lambda$CDM model, and also shows a steep decay.
\end{abstract}


\section{Introduction}
The star formation history (SFH), especially at high-redshift
($z>6$), is important in many fields in astrophysics. Direct SFR
measurements are quite challenging at high-redshift, particularly at
the faint end of the galaxy luminosity function. The star formation
rate (SFR) has been comprehensively investigated. Hopkins \& Beacom
(2006) calibrated the star formation history out to $z>6$ using the
ultraviolet and far-infrared data. They found the SFR is tightly
constrained at $z<1$, the SFR in the redshift range of $1<z<4$ is
approximately a constant and shows a steep decay at $z>4$ (Hopkins
\& Beacom2006). Li (2008) constrained the SFR up to $z=7.4$ by
adding 6 new data and found the same result as Hopkins \& Beacom
(2006)(Li 2008). The high-redshift SFR can also be determined by
observations of color-selected Lyman break galaxies (LBGs) (Bouwens
et al. 2008; Mannucci et al. 2007; Verma et al. 2007). Ota et al.
(2008) constrained the SFR using Ly$\alpha$ emitters (LAEs)(Ota et
al. 2008). In Figure 1, we list the different observational results.
We can see that different results disagree with each other even
considering the uncertainties.

Long-duration gamma-ray bursts triggered by the death of massive
stars, which have been shown to be associated with supernovae
(Stanek et al. 2003; Hjorth et al. 2003), provide a complementary
technique for measuring the SFR. GRBs at high redshfits are
predicated to be observed  because of their high luminosities (Lamb
\& Reichart 2000; Ciardi \& Loeb 2000;  Bromm \& Loeb 2002, 2006;
Gou et al. 2004). The farthest GRB to date is GRB 090423 at $z=8.3$
(Tanvir et al. 2009; Salvaterra et al. 2009). So GRBs are a
promising probe of the star formation history (Totani 1997; Wijers
et al. 1998; Porciani \& Madau 2001; Bromm \& Loeb 2002,2006;
Tutukov 2003). Recent studies show that \emph{Swift} GRBs are not
tracing the star formation history exactly but including an
additional evolution (Daigne et al. 2006; Cen \& Fang 2007; Le \&
Dermer 2007; Y\"{u}ksel \& Kistler 2007; Salvaterra \& Chincarini
2007a; Guetta \& Piran 2007; Kistler et al. 2008; Salvaterra et al.
2008; Butler et al. 2009). Observations show that GRBs prefer to
form in a low-metallicity environment (Le et al. 2003; Stanek et al.
2006). In the collapsar paradigm, large angular momentum is required
to power a GRB. Because high-metallicity stars are expected to have
significant mass loss through winds promoting the loss of angular
momentum, Langer \& Norman (2006) and Woosley \& Heger (2006) have
argued that GRB progenitors will have a low metallicity (Woosley \&
Heger 2006; M\'{e}sz\'{a}ros 2006; Langer \& Norman 2006). This has
implications for the expected redshift distribution of GRBs
(Natarajan et al. 2005; Salvaterra \& Chincarini 2007a; Salvaterra
et al. 2007b; Li 2008). Chary et al. (2007) estimated a lower limit
of the SFR of $0.12\pm 0.09$ and $0.09\pm 0.05$ $\rm M_{\odot}\rm
yr^{-1}\rm Mpc^{-3}$ at $z=4.5$ and $6$, respectively, using deep
observations of three $z\sim 5$ GRBs with the \emph{Spitzer Space
Telescope}(Chary, Berger \& Cowie 2007). Y\"{u}ksel et al.(2008)
used \emph{Swift} GRB data to constrain the high-redshift SFR and
found that no steep drop exists in the SFR up $z\sim 6.5$
(Y\"{u}ksel et al. 2008). Kistler et al. (2009) constrained SFR up
to $z\sim8$ using four-years Swift data and found SFR to beyond $z =
8$ was consistent with LBG-based measurements (Kistler et al. 2009).

In this paper, we estimate the high-redshfit SFR using latest
\emph{Swift} long-duration GRBs, considering the GRB formation rate
tracing SFH and the cosmic metallicity evolution. We use the SFR
between $z=1$ and $z=4$ and relate the GRB counts in this redshift
bin. The absolute conversion factor between the SFR and the GRB rate
is highly uncertain. But we do not need this factor in this method.
Because weak low-redshift GRBs can not be seen at high redshifts, so
we only use high luminosity GRBs. We proceed analogously to
Y\"{u}ksel et al. (2008). But there are two differences between our
method and Y\"{u}ksel's method. First, we consider that long GRBs
prefer to form in low-metallicity regions and trace the star
formation history, but Y\"{u}ksel et al. (2008) and Kistler et al.
(2009) considered GRBs do not trace star formation history directly,
instead implying some kind of additional evolution. Second, we
examine the influence of background cosmology. In 1998, observations
on type Ia supernovae suggest the accelerating universe. Many models
have been proposed to explain the accelerating expansion. Out of
many particular models, we focus on four representative models:
$\Lambda$CDM, quintessence, quintessence with time-varying equation
of state, and brane-world.

\begin{figure}
\includegraphics[width=\linewidth]{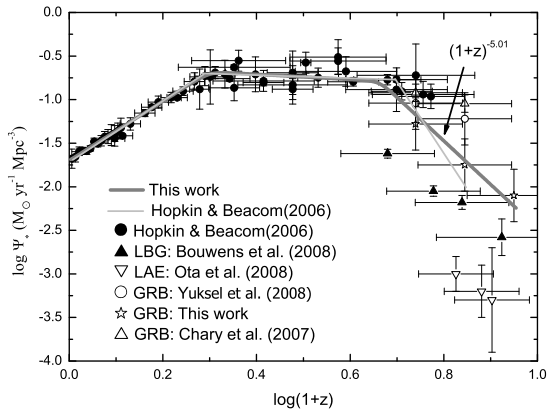}
\caption{\label{Fig1} The cosmic star formation history. The
 black circles are from Hopkins \& Beacom (2006). The three open pentacles
 are the star formation rates at $z=4.5$, $z=6$ and $z=8$ derived using \emph{Swift} GRB data.}
\end{figure}

The structure of this paper is as follows. In section 2, we
introduce the method. In section 3, we show our results on
high-redshift SFR. Finally, section 4 contains conclusions and
discussions.

\section{The method}
More recent observational studies indicated that the long GRB host
galaxy metallicity is generally lower than that of the average
massive star forming galaxies (Le et al. 2003; Stanek et al. 2006).
Salvaterra \& Chincarini (2007) found that the differential peak
flux number counts obtained by BATSE and \emph{Swift} could be well
fitted using GRBs forming in low-metallicity galaxies (Salvaterra \&
Chincarini 2007a). Under the assumption that the formation of GRBs
follows the cosmic star formation history and GRBs form
preferentially in low-metallicity galaxies, the GRB formation rate
is given by
\begin{equation}
\Psi_{\rm GRB}(z)=k_{\rm GRB}\Sigma (Z_{th},z)\Psi _*(z),
\end{equation}
where $k_{\rm GRB}$ is the GRB formation efficiency, $\Sigma
(Z_{th},z)$ is the fraction of galaxies at redshift $z$ with
metallicity below $Z_{\rm th}$ (Langer \& Norman 2006) and $\Psi
_*(z)$ is the observed comoving SFR. The redshift distribution of
observable GRBs is
\begin{equation}
\frac{d N}{d z}=F(z)/\langle f_{\rm beam}\rangle \Sigma
(Z_{th},z)\Psi _*(z)\frac{dV/dz}{1+z},
\end{equation}
where $F(z)$ represents the ability both to detect the trigger of
burst and to obtain the redshift (Kistler et al. 2008). The
redshifts of high-redshift GRBs are determined by a spectral break
in near infrared or infrared bands. Many ground-based facilities
could recognize the spectral break and then obtain the redshift.
GROND observed the spectral break of GRB080913 between i$'$ and z$'$
bands and the redshift of GRB080913 is z=6.7 (Greiner et al. 2009).
Greiner et al. (2009) show that 2m-class telescopes can identify
most high-redshift GRBs. The redshift of GRB090423 is determined by
NIR spectroscopic observations (Tanvir et al. 2009; Salvaterra et
al. 2009). So if the luminosities of high-redshift GRBs are high
enough and spacecrafts (such as Swift and Fermi) can detect, the
redshifts can be obtained using ground-based facilities. $\langle
f_{\rm beam}\rangle$ is the beaming factor and ${\rm d}V/{\rm d}z$
is the comoving volume element per unit redshift, given by
\begin{equation}
\frac{{\rm d}V}{{\rm d}z}=\frac{4\pi c d_{L}^{2}}{1+z}
\frac{H(z)}{1+z}.
\end{equation}
The luminosity distance, $d_{L}$, to a source at redshift $z$ is
\begin{equation}
d_{L}=c(1+z)\int_{0}^{z}\frac{1}{H(z')} {\rm d}z' ,
\end{equation}
with
\begin{equation}
H(z)=H_0\sqrt{\Omega_{m}(1+z)^{3}+\Omega_{\Lambda}}
\end{equation}
in a flat $\Lambda$CDM universe. For bursts with luminosities
sufficient to be viewed within an entire redshift range, Kistler et
al. (2008) found that $F(z)$ could be set to a constant (for more
details, see Kistler et al. 2008, 2009).

Some theoretical models (Woosley \& Heger 2006; see M\'{e}sz\'{a}ros
2006 for a review) require that GRB progenitors should have
metallicity $\leq0.1 Z_{\odot}$. According to Langer \& Norman
(2006), the fractional mass density belonging to metallicity below a
given threshold $Z_{\rm th}$ is (Langer \& Norman 2006)
\begin{equation}
\Sigma (Z_{\rm th},z)=\frac{\hat{\Gamma}[\alpha_1+2,(Z_{\rm
th}/Z_{\odot})^2 10^{0.15\beta z}]}{\Gamma(\alpha_1+2)},
\end{equation}
where $\hat{\Gamma}$ and $\Gamma$ are the incomplete and complete
gamma functions, $\alpha_1=-1.16$ is the power-law index in the
Schechter distribution function of galaxy stellar masses (Panter,
Heavens \& Jimenez 2004) and $\beta=2$ is the slope  of the galaxy
stellar mass-metallicity relation (Savaglio et al. 2005; Langer \&
Norman 2006). We adopt $Z_{\rm th}=0.1Z_{\odot}$ as in Langer \&
Norman (2006).

We show the luminosity-redshift distribution of 119 long GRBs
observed by \emph{Swift}\footnote{See $\rm
http://swift.gsfc.nasa.gov/docs/swift/archive/grb\_table.$} till GRB
090529 in Figure 2. The isotropic luminosity is
\begin{equation}
L_{\rm iso}=E_{\rm iso}/[T_{\rm 90}/(1+z)],
\end{equation}
where $E_{\rm iso}$ is the isotropic energy in the $1-10^4$keV band
and $T_{\rm 90}$ is the GRB duration. Because only bright bursts can
be seen at low and high redshifts, so we choose the luminosity cut
$L_{\rm iso}>3\times 10^{51}$ ergs $s^{-1}$ (Y\"{u}ksel et al.
2008). There are four groups of GRBs defined by this $L_{\rm iso}$
cut in $z \!=\!$ $1-4$, $4-5$, $5-7$, and $7-8.5$.  The GRBs in $z =
1-4$ act as a ``control group'' to base the GRB to SFR conversion,
since this redshift bin has both good SFR measurements and GRB
counts. We calculate the theoretically predicated number of GRBs in
this bin as
\begin{eqnarray}
N_{1-4}^{\rm the} & = & \Delta t \frac{\Delta \Omega}{4\pi}
\int_{1}^{4} dz\,  F(z) \, \Sigma (Z_{\rm th},z) \frac{\Psi _*(z)}
{\left\langle f_{\rm beam}\right\rangle} \frac{dV/dz}{1+z} \nonumber \\
& = & A \, \int_{1}^{4} dz\, \Sigma (Z_{\rm th},z)\, \Psi _*(z) \,
\frac{dV/dz}{1+z}\,, \label{N1-4}
\end{eqnarray}
where $A = {\Delta t \, \Delta \Omega \, F_0} / 4\pi {\left\langle
f_{\rm beam} \right\rangle}$ depends on the total time, $\Delta t$,
and the angular sky coverage, $\Delta \Omega$. The theoretical
number in $z = 4-5$ and $5-7$ can be written by
\begin{eqnarray}
N_{z_1-z_2}^{\rm the} & = &  \left\langle \Psi _*
\right\rangle_{z_1-z_2} A \, \int_{z_1}^{z_2} dz\, \Sigma (Z_{\rm
th},z) \, \frac{dV/dz}{1+z}\,, \label{Nz1-z2}
\end{eqnarray}
where $\left\langle \dot{\rho}_* \right\rangle_{z_1-z_2}$ is the
average SFR density in the redshift range $z_1-z_2$.  Representing
the predicated numbers, $N_{z_1-z_2}^{\rm the}$ with the observed
GRB counts, $N_{z_1-z_2}^{\rm obs}$, we obtain the SFR in the
redshift range $z_1-z_2$
\begin{equation}
\left\langle \Psi_* \right\rangle_{z_1-z_2} = \frac{N_{z_1-z_2}^{\rm
obs}}{N_{1-4}^{\rm obs}} \frac{\int_{1}^{4} dz\,
\frac{dV/dz}{1+z}\Sigma (Z_{\rm th},z) \Psi_*(z)\,
}{\int_{z_1}^{z_2} dz\, \frac{dV/dz}{1+z}\Sigma (Z_{\rm th},z)}\,.
\label{zratio}
\end{equation}

\begin{figure}
\includegraphics[width=\linewidth]{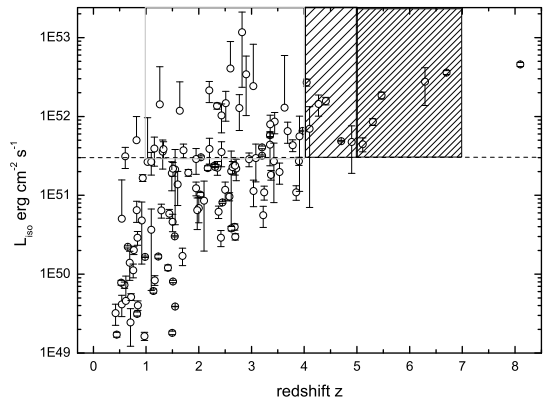}
 \caption{\label{Fig2} The luminosity-redshift distribution of 119 \emph{Swift} long duration GRBs in the $\Lambda$CDM model.
 The number counts in redshift bins $z=1-4$, $4-5$, $5-7$, and $7-8.5$ are $28$, $6$, $5$, and $1$ respectively.}
\end{figure}

Below, we briefly introduce three other cosmological models in which
we calculate the high-redshift SFR. We will restrict our attention
to flat models ($k = 0$) because the flat geometry is strongly
supported by five-years WMAP data (Komatsu et al. 2009).

We first consider the dark energy with a constant equation of state
$w(z)=w_0$, where $-1<w<-1/3$. In such a case this component is
called ``quintessence". Confrontation with recent observational
datasets, Wang et al. (2007) found $\Omega_M=0.31$ and $w_0=-0.95$.
We use these data in the following calculations (Wang, Dai \& Zhu
2007). The luminosity-redshift distribution of GRBs in this model is
shown in Figure 3.

\begin{figure}
\includegraphics[width=\linewidth]{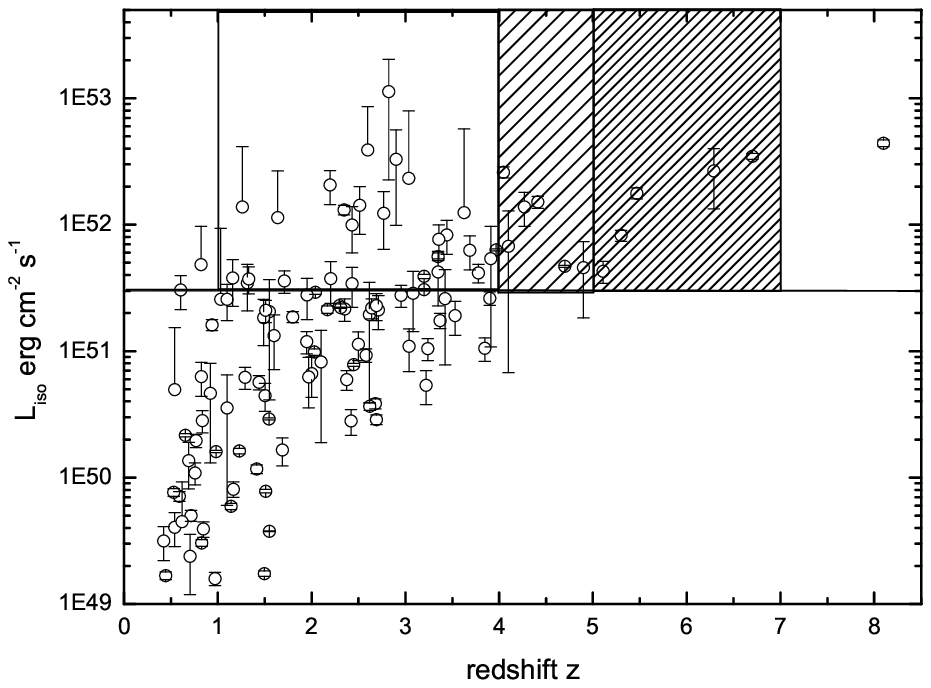}
 \caption{\label{Fig3} The luminosity-redshift distribution of 119 \emph{Swift} long duration GRBs in the quintessence ($w=w_0$).
 The number counts in redshift bins $z=1-4$, $4-5$, $5-7$, and $7-8.5$ are $27$, $6$, $5$, and $1$ respectively. }
\end{figure}

If we consider that the quintessence arises from an evolving scalar
field, it would be natural to expect that the equation of state
should vary with time. We consider $w(z)=w_0+w_1z/(1+z)$. We also
use the results of Wang et al. (2007), $\Omega_M=0.30$, $w_0=-1.08$
and $w_1=0.84$ (Wang, Dai \& Zhu 2007). The luminosity-redshift
distribution of GRBs in this model is shown in Figure 4.

\begin{figure}
\includegraphics[width=\linewidth]{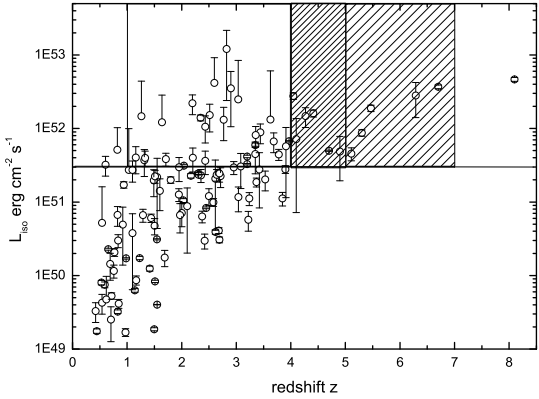}
 \caption{\label{Fig3} The luminosity-redshift distribution of 119 \emph{Swift} long duration GRBs in the quintessence with
 a time-varying equation of state ($w=w_0+w_1z/(1+z)$).
 The number counts in redshift bins $z=1-4$, $4-5$, $5-7$, and $7-8.5$ are $30$, $6$, $5$, and $1$ respectively. }
\end{figure}

Brane-world scenarios assume that our four-dimensional space-time is
embedded into five-dimensional space (Deffayet, Davli \& Gabadadze
2002). Gravity in five dimensions is governed by the usual
five-dimensional Einstein-Hilbert action. The bulk metric induces a
four-dimensional metric on the brane. We consider the flat
Dvali-Gabadadze-Porrati model. The only parameter is $\Omega_M$. We
use the result from Wang et al. (2009), $\Omega_M=0.20$ (Wang, Dai
\& Qi 2009). The luminosity-redshift distribution of GRBs in this
model is shown in Figure 5.

\begin{figure}
\includegraphics[width=\linewidth]{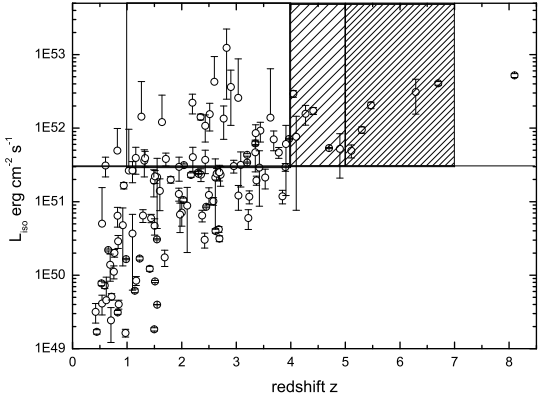}
\caption{\label{Fig4} The luminosity-redshift distribution of 119
\emph{Swift} long duration GRBs in the brane-world model. The number
counts in redshift bins $z=1-4$, $4-5$, $5-7$, and $7-8.5$ are $31$,
$6$, $5$, and $1$ respectively.}
\end{figure}

The SFR conversion from one cosmology to another is as follows. In
the flat universe, the comoving volume is proportional to comoving
distance cubed, $V_c\propto D_c^3$, and the comoving volume between
redshifts $z-\Delta z$ and $z+\Delta z$ is $V_c(z,\Delta z)\propto
D_c^3(z+\Delta z)-D_c^3(z-\Delta z)$. Since the luminosity is
proportional to the comoving distance squared, $L\propto D_c^2$, the
SFR density for a given redshift range is (Hopkins 2004)
\begin{equation}
\Psi _*(z)\propto\frac{L(z)}{V_c(z,\Delta z)}\propto
\frac{D_c^2(z)}{D_c^3(z+\Delta z)-D_c^3(z-\Delta z)}.
\end{equation}
The Hubble functions in different dark energy models are showed in
Table 1. We convert the SFR in the redshift range $z=1-4$ in
$\Lambda$CDM to other models. The derived SFR can be used to
calculate the high-redshift SFR in quintessence, quintessence with a
time-varying equation of state and brane-world model.

\begin{table*}
\begin{minipage}{10cm}
\caption{Expansion rates $H(z)$ in four models. We consider the flat
universe. In the brane-world model,
$\Omega_{r_c}=(1-\Omega_m^2)/4$.}
\bigskip
 \begin{tabular}{|c|c|} \hline
 Model & Cosmological expansion rate $H(z)$). \\
\hline
  $\Lambda$CDM & $H^2(z) = H^2_0 \left[ \Omega_m \; (1+z)^3 + 1-\Omega_m \right]$  \\
Quintessence & $H^2(z) = H^2_0 \left[ \Omega_m \; (1+z)^3 +
(1-\Omega_m) \; (1+z)^{3(1+w)} \right]
$ \\
Var Quintessence & $H^2(z) = H^2_0 \left[ \Omega_m \; (1+z)^3 +
(1-\Omega_m) \;
(1+z)^{3(1+w_0-w_1)}\;\exp(3 w_1 z) \right]$  \\
Braneworld & $H^2(z) = H_0^2 \left[ (\sqrt{ \Omega_{m} (1+z)^3 +
\Omega_{r_c} }
+ \sqrt{\Omega_{r_c}} )^2 \right]$   \\
\hline
\end{tabular}
\end{minipage}
\end{table*}

\section{The derived high-redshift star formation rate}
In Figure 1, we show the measurement of the high-redshift SFR in the
$\Lambda$CDM model with $\Omega_M=0.27$, $\Omega_{\Lambda}=0.73$ and
$H_0=70 {\rm km}\,{\rm s}^{-1}\,{\rm Mpc}^{-1}$. The star formation
rates are Log$\Psi _*=-1.287\pm 0.30$, Log$\Psi _*=-1.750\pm 0.30$,
and Log$\Psi _*=-2.110\pm 0.30$ $\rm M_{\odot}\rm yr^{-1}\rm
Mpc^{-3}$ at $z=4.5$, $6.0$, and $8.0$, respectively. Taking into
account the Poisson confidence interval for four observed GRBs, we
assign a statistical uncertainty of a factor of 2 (Y\"{u}ksel et al.
2008). The derived high-redshift SFR shows a steep decay with a
slope of about $5.0$ (the slope of Hopkins \& Beacom (2006) is about
$7.8$). If we exclude a particular GRB or changes in redshift
ranges, the derived SFR will change insignificantly. This conclusion
is consistent with Y\"{u}ksel et al. (2008), but the derived
high-redshift SFR is different from that of Y\"{u}ksel et al.
(2008). The main reason is that we consider GRBs prefer to form in
low-metallicity regions.

We show the update SFH fit of Hopkins \& Beacom (2006) at
high-redshifts based on our new GRB results. We use a continuous
form of a broken power law (Y\"{u}ksel et al. 2008),
\begin{eqnarray}
\Psi_*(z) & = & \Psi_{*,0}  \left\{ \,
\left[ (1 + z)^{a}\right]^{\eta}+\left[ \frac{(1 + z)^{b}}{(1 + z_1)^{b-a} }\right]^{\eta}\right. \nonumber \\
&  & \left. +\left[ \frac{(1 + z)^{c}}{ (1 + z_1)^{b-a} (1 +
z_2)^{c-b} }\right]^{\eta} \,\right\}^{1/\eta}\,,
\end{eqnarray}
where using $\eta \simeq-10$ smoothes the power law transitions. Our
fitted result is shown by the thick gray line in Fig 1.  Here, $z_1=
1$ and $z_2 = 4.0$ are the redshifts of the breaks, $a = 3.4$, $b =
-0.3$ and $c = -5.0$ is the slopes of these three power laws, and
the normalization is $\Psi_{*,0} = 0.02
\,M_\odot$~yr$^{-1}$~Mpc$^{-3}$.

\begin{table}
\caption{The high-redshift SFRs in four models.}
\bigskip
 \begin{tabular}{|c|c|c|} \hline
 Model & Redshift & Star formation rate \\
\hline
  $\Lambda$CDM & $4.5\pm 0.5$ & Log$\Psi _*=-1.287\pm 0.30$ \\
  $\Lambda$CDM & $6.0\pm 1.0$ & Log$\Psi _*=-1.750\pm 0.30$ \\
  $\Lambda$CDM & $8.0\pm 0.5$ & Log$\Psi _*=-2.110\pm 0.30$ \\
  Quintessence & $4.5\pm 0.5$ & Log$\Psi _*=-1.296\pm 0.30$ \\
  Quintessence & $6.0\pm 1.0$ & Log$\Psi _*=-1.787\pm 0.30$ \\
  Quintessence & $8.0\pm 0.5$ & Log$\Psi _*=-2.287\pm 0.30$ \\
Var Quintessence & $4.5\pm 0.5$ & Log$\Psi _*=-1.394\pm 0.30$ \\
Var Quintessence & $6.0\pm 1.0$ & Log$\Psi _*=-1.819\pm 0.30$ \\
Var Quintessence & $8.0\pm 0.5$ & Log$\Psi _*=-2.214\pm 0.30$ \\
Brane-world & $4.5\pm 0.5$ & Log$\Psi _*=-1.466\pm 0.30$ \\
Brane-world & $6.0\pm 1.0$ & Log$\Psi _*=-1.905\pm 0.30$ \\
Brane-world & $8.0\pm 0.5$ & Log$\Psi _*=-2.205\pm 0.30$ \\
\hline
\end{tabular}

\end{table}

In Table 2, we show the high-redshift SFRs in four cosmological
models. The value of SFR in these models show no remarkable
difference. The reasons are as follows. First, the GRB counts in
redshift bins are different because the luminosity distances in
these models are not equal. Second, the comoving volume of these
models are different. Third, there is a conversion factor from one
cosmological model to another. The derived SFR in these cosmological
models are consistent with each other in quoted uncertainties. The
influence of background cosmology can be neglected when we use this
method to measure the high-redshift SFR.

\section{conclusions and discussions}
In this paper, we constrain the high-redshift SFR up to $z\sim8.3$
using the latest \emph{Swift} GRB data including two highest-$z$
GRBs, GRB 080913 at $z=6.7$ and GRB 090423 at $z=8.3$ in four
background cosmological models. We consider that GRBs trace the star
formation history and prefer to form in a low-metallicity
environment. The empirical method is similar to Y\"{u}ksel et al.
(2008) and Kistler et al. (2009). This method has two advantages.
First, statistics of the recent \emph{Swift} GRB data allowed the
use of luminosity cuts to fairly compare GRBs in the full redshift
range, eliminating the unknown GRB luminosity function. Second, we
can calculate the high-redshift SFR by comparing the counts of GRBs
at different redshift ranges, normalized to SFR data at intermediate
redshifts which have been well constrained by observations,
eliminating the need for knowledge of the GRB efficiency factor. But
there are two differences between our method and Y\"{u}ksel's
method. First, we consider that long GRBs prefer to form in
low-metallicity regions and trace the star formation history, but
Y\"{u}ksel et al. (2008) and Kistler et al. (2009) considered GRBs
do not trace star formation history directly, instead implying some
kind of additional evolution. Second, we examine the influence of
background cosmology.

Our results show that the SFR at $z>4$ shows a steep decay with a
slope of about $-5.0$ in $\Lambda$CDM. In three other models, the
high-redshift SFR is slightly different from that in the
$\Lambda$CDM model, and also shows a steep decay. Y\"{u}ksel et al.
(2008) found the decay slope was $-3.5$ at $z>4$. Our derived
high-redshift SFR is different from Y\"{u}ksel et al. (2008). The
main reason is that we consider GRBs prefer to form in
low-metallicity regions. Li (2008) derived the SFH up to $z=7.4$,
and found that the decay slope at $z>4$ was $4.48$ (Li 2008). Their
result is consistent with ours.

\section*{ACKNOWLEDGMENTS}
We thank A. M. Hopkins and H. Y\"{u}ksel for sharing the SFR data.
This work was supported by the National Natural Science Foundation
of China (grants 10233010, 10221001 and 10873009) and the National
Basic Research Program of China (973 program) No. 2007CB815404. F.
Y. Wang was also supported by Jiangsu Project Innovation for PhD
Candidates (CX07B-039z).

\end{document}